\documentclass{kluwer_mod}    % Specifies the document style.
\usepackage{graphicx}
\usepackage{klucite}

\def\figref#1{figure~\ref{#1}}
\def\Figref#1{Figure~\ref{#1}}

\def\Re{{\rm Re}}
\def\Rel{{\rm Re}_\lambda}
\def\Pr{{\rm Pr}}

\def\lf{\ell_0}
\def\ld{\ell_\nu}
\def\lres{\ell_\eta}
\def\ls{\ell_{\rm s}}

\def\kd{k_\nu}
\def\kl{k_\lambda}
\def\kres{k_\eta}
\def\krms{k_{\rm rms}}
\def\kpar{k_\parallel}

\def\vk{{\bf k}}
\def\vu{{\bf u}}
\def\vB{{\bf B}}

\def\vb{{\skew{-4}\hat{\bf b}}}

\def\<{\langle}
\def\>{\rangle}

\def\usq{{\langle u^2 \rangle}}
\def\Bsq{{\langle B^2 \rangle}}
\def\Bfr{{\langle B^4 \rangle}}

\begin{document}                                    
\begin{article}

\begin{opening}         
\title{From Small-Scale Dynamo to Isotropic MHD Turbulence} 
\author{Alexander~A.~\surname{Schekochihin}}
\author{Steven~C.~\surname{Cowley}}
\author{Samuel~F.~\surname{Taylor}}
\institute{Imperial College, London, U.K.; E-mail: as629\@damtp.cam.ac.uk}
\author{Jason~L.~Maron}
\institute{University of Rochester, Rochester, NY, U.S.A.}
\author{James~C.~McWilliams}
\institute{UCLA, Los Angeles, CA, U.S.A.}
\runningauthor{A.~A.~Schekochihin et al.}
\runningtitle{From Small-Scale Dynamo to Isotropic MHD Turbulence}
\date{\today}

\begin{abstract}
We consider the problem of incompressible, forced, nonhelical, homogeneous, 
isotropic MHD turbulence with no mean magnetic field. 
This problem is essentially different from the case with externally 
imposed uniform mean field. There is no scale-by-scale 
equipartition between magnetic and kinetic energies as would be the case 
for the Alfv\'en-wave turbulence. The isotropic MHD turbulence is the 
end state of the turbulent dynamo which generates folded 
fields with small-scale direction reversals. We propose that the 
statistics seen in numerical simulations of isotropic MHD turbulence 
could be explained as a superposition of these folded fields and 
Alfv\'en-like waves that propagate along the folds.  
\end{abstract}

\end{opening}
           
The term 'MHD turbulence' embraces a number of turbulent regimes 
described by the MHD equations. Their physics can be very different 
depending on the Mach number, presence of external forcing 
and of mean magnetic field, flow helicity, 
relative magnitude of the velocity and magnetic-field diffusion 
coefficients, etc. Here we consider what is perhaps 
the oldest MHD turbulence problem dating back 
to \inlinecite{Batchelor_dynamo}: 
incompressible, randomly forced, nonhelical, homogeneous, isotropic 
MHD turbulence. No mean field is imposed, 
so all magnetic fields are fluctuations 
generated by the turbulent dynamo. We are primarily interested 
in the case of large magnetic Prandtl number~$\Pr=\nu/\eta$ (the ratio 
of fluid viscosity to magnetic diffusivity), which is appropriate  
for the warm ISM and cluster plasmas. Numerical 
evidence suggests that the popular choice~$\Pr=1$ is 
in many ways similar to the large-$\Pr$ regime. $\Pr\gg1$ implies 
that the resistive scale~$\lres\sim\Pr^{-1/2}\ld$ is much smaller than 
the viscous scale~$\ld$. 
Thus, the problem has two scale ranges: the hydrodynamic (Kolmogorov) 
inertial range $\lf\gg\ell\gg\ld\sim\Re^{-3/4}\lf$ 
($\lf$ is the forcing scale) 
and the subviscous range $\ld\gg\ell\gg\lres$. 
This makes it very hard to simulate this regime numerically.  

For a moment, let us consider the traditional view of the fully developed 
incompressible MHD turbulence in the presence of a strong externally 
imposed mean field. This view is based on the idea of 
\inlinecite{Iroshnikov} and 
\inlinecite{Kraichnan_MHD} that it is a turbulence 
of strongly interacting Alfv\'en-wave packets. 
Their phenomenology, modified by 
\inlinecite{GS_strong} to account for the anisotropy induced by 
the mean field, predicts steady-state spectra for magnetic 
and kinetic energies that are identical in the inertial range and 
have Kolmogorov $k^{-5/3}$ scaling. 
An essential feature of this 
description is that {\em it implies scale-by-scale 
equipartition between magnetic and velocity fields:} indeed, 
$\delta\vu_\vk=\delta\vB_\vk$ in an Alfv\'en wave. 
%The natural 
%description of such a turbulence is in terms of Els\"asser variables 
%$\vz^\pm=\vu\pm\delta\vB$ (here $\delta\vB=\vB-\vB_0$).
Numerics appear to confirm the Alfv\'enic equipartition  
picture {\em provided there is an externally imposed strong 
mean field.} The reader will find further details and references 
in Chandran's review in these Proceedings. 

\begin{figure}[t]
\centerline{\includegraphics[width=2.4in]{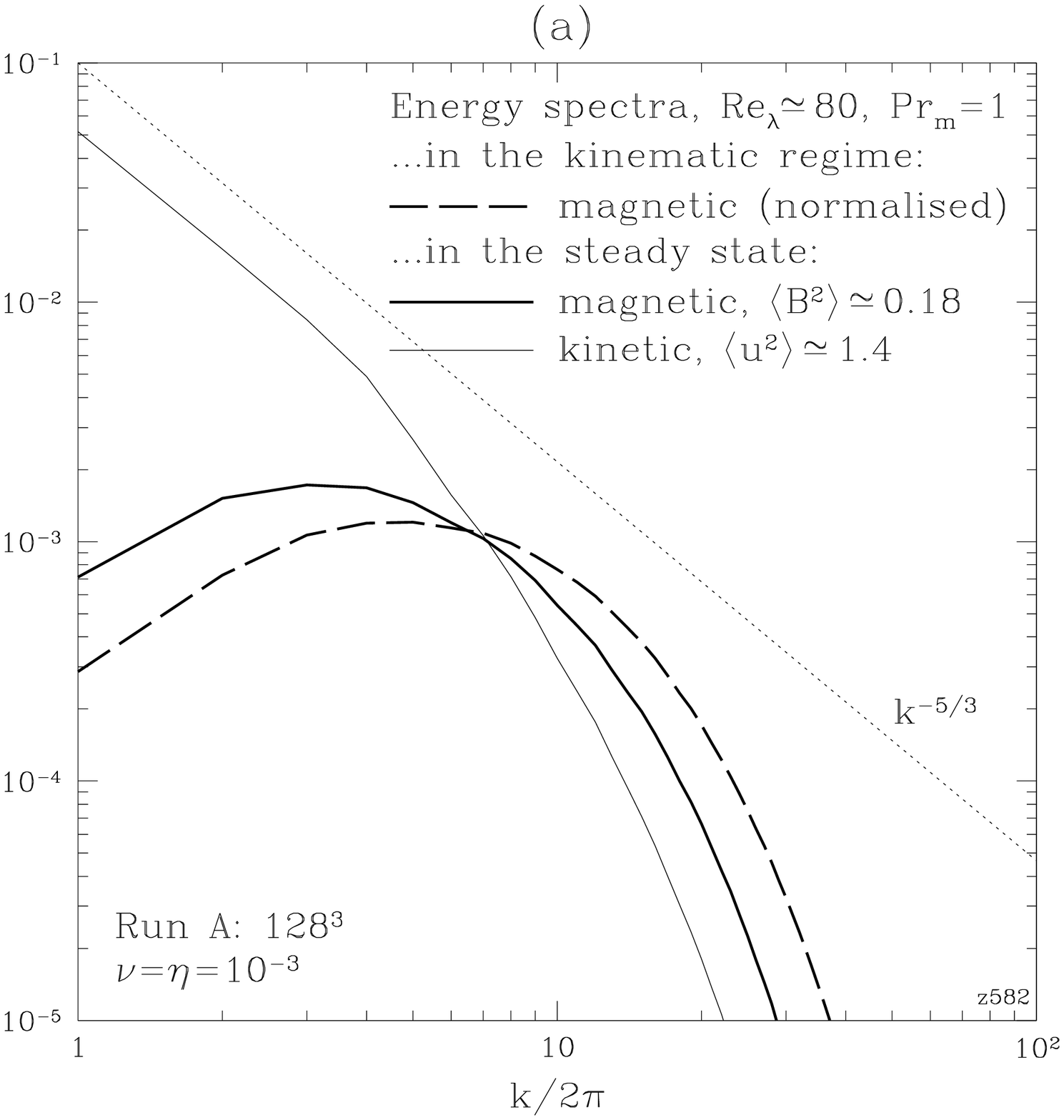}
\includegraphics[width=2.4in]{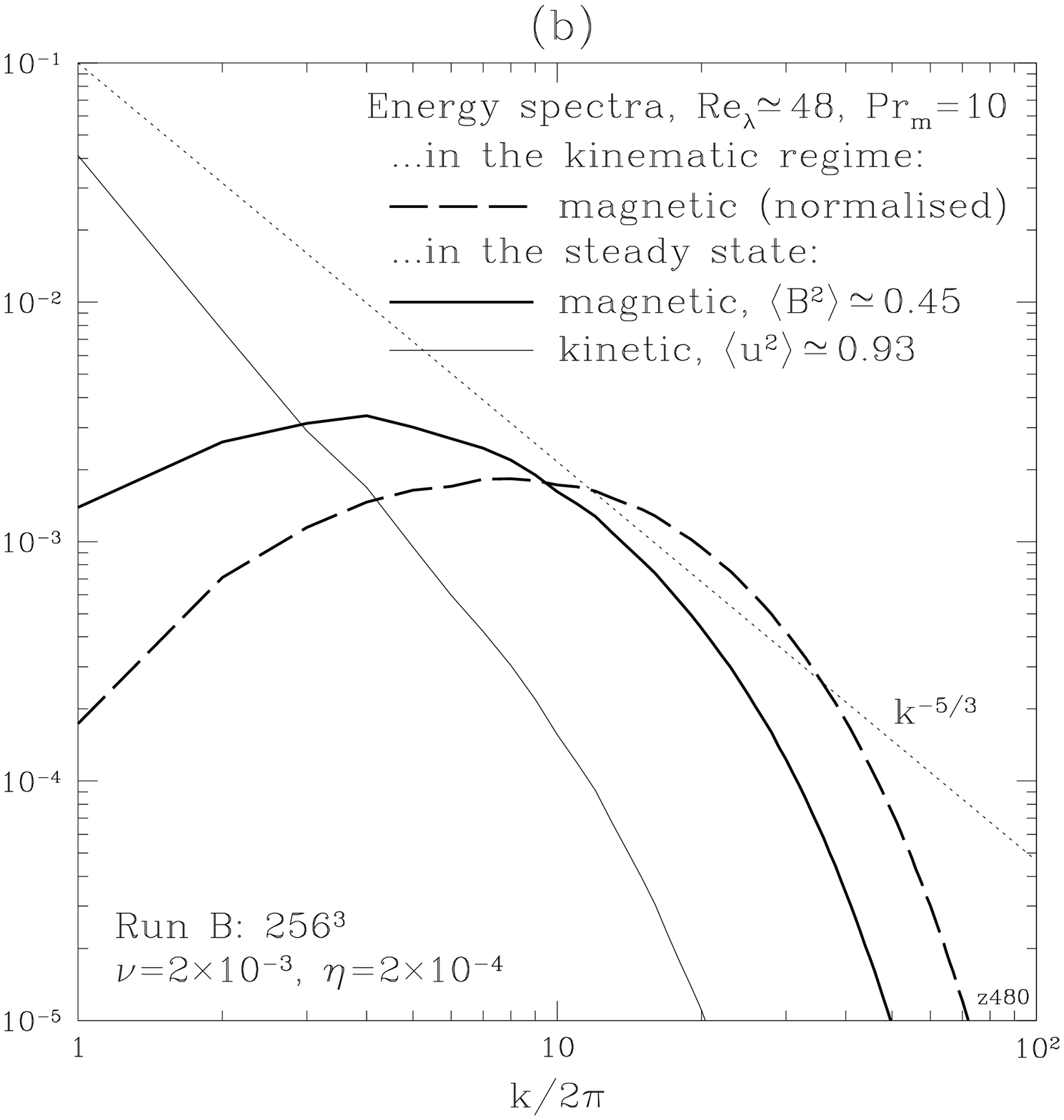}}
\caption{Energy spectra in our simulations of isotropic MHD 
turbulence. Here $\Rel=\usq^{1/2}\lambda/\nu\sim\Re^{1/2}$, where  
$\lambda=\sqrt{5}\,(\<|\nabla\vu|^2\>/\usq)^{-1/2}$ 
is the Taylor microscale and $\Re$ the box Reynolds number.}\label{fig_spectra}
\end{figure}

In the case of zero mean field, it is tempting to argue that essentially 
the same description applies, except now it is 
the large-scale magnetic fluctuations that 
play the role of effective mean field along which smaller-scale Alfv\'en waves 
can propagate. This is, indeed, what has widely been assumed to be true. 
However, 
the numerical simulations of isotropic MHD turbulence tell a very different story. 
{\em No scale-by-scale equipartition between kinetic and magnetic energies is 
observed numerically.} There is a definite and very significant excess of 
magnetic energy at small scales. This holds true both for $\Pr>1$ and for $\Pr=1$ 
(\figref{fig_spectra}). The trend is evident already at 
low resolutions and persists at the highest currently available 
resolution ($1024^3$, see \opencite{Haugen_Brandenburg_Dobler}
and their paper in these Proceedings).

\begin{figure}[t]
\centerline{\includegraphics[width=2.4in]{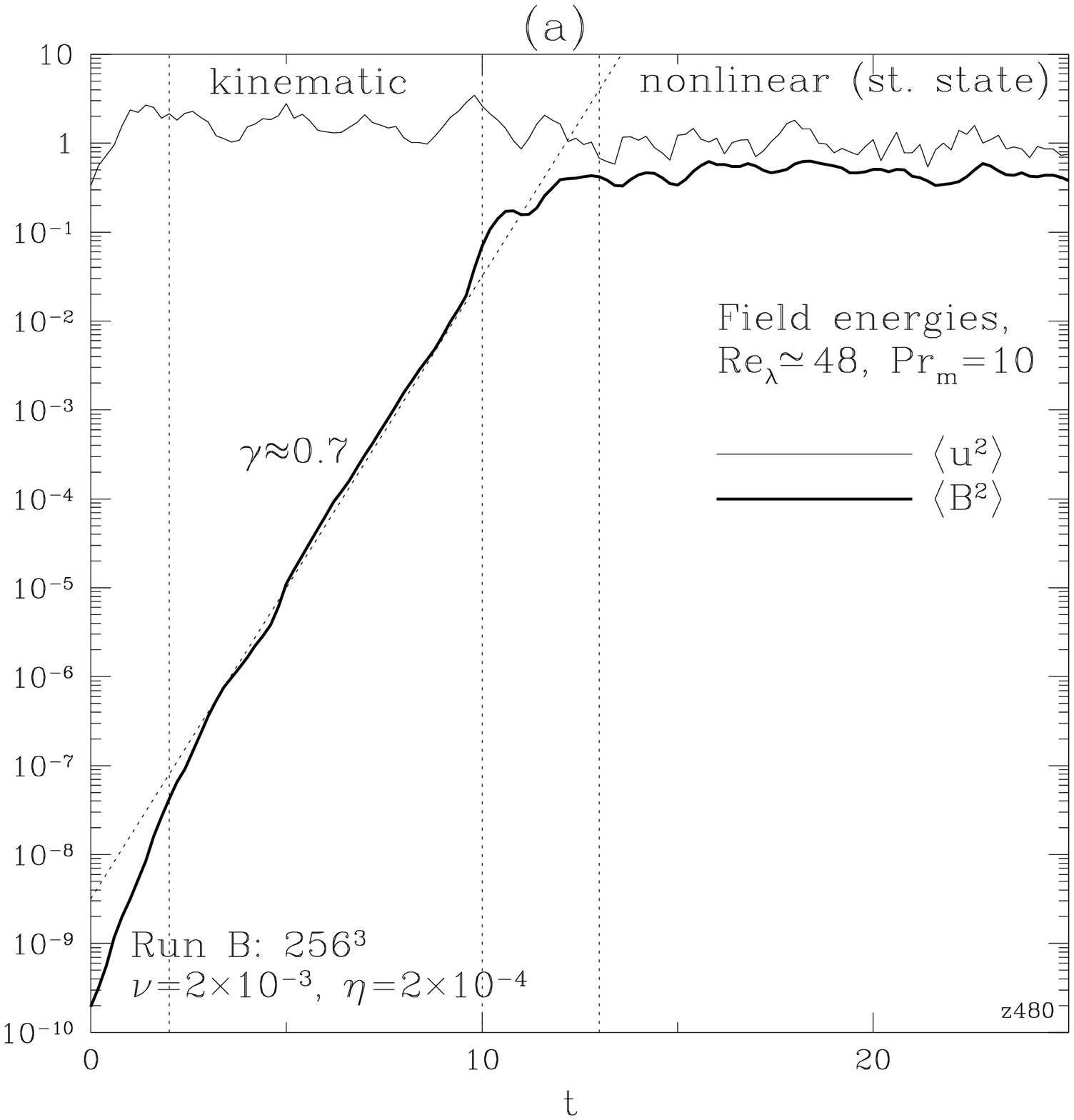}
\includegraphics[width=2.4in]{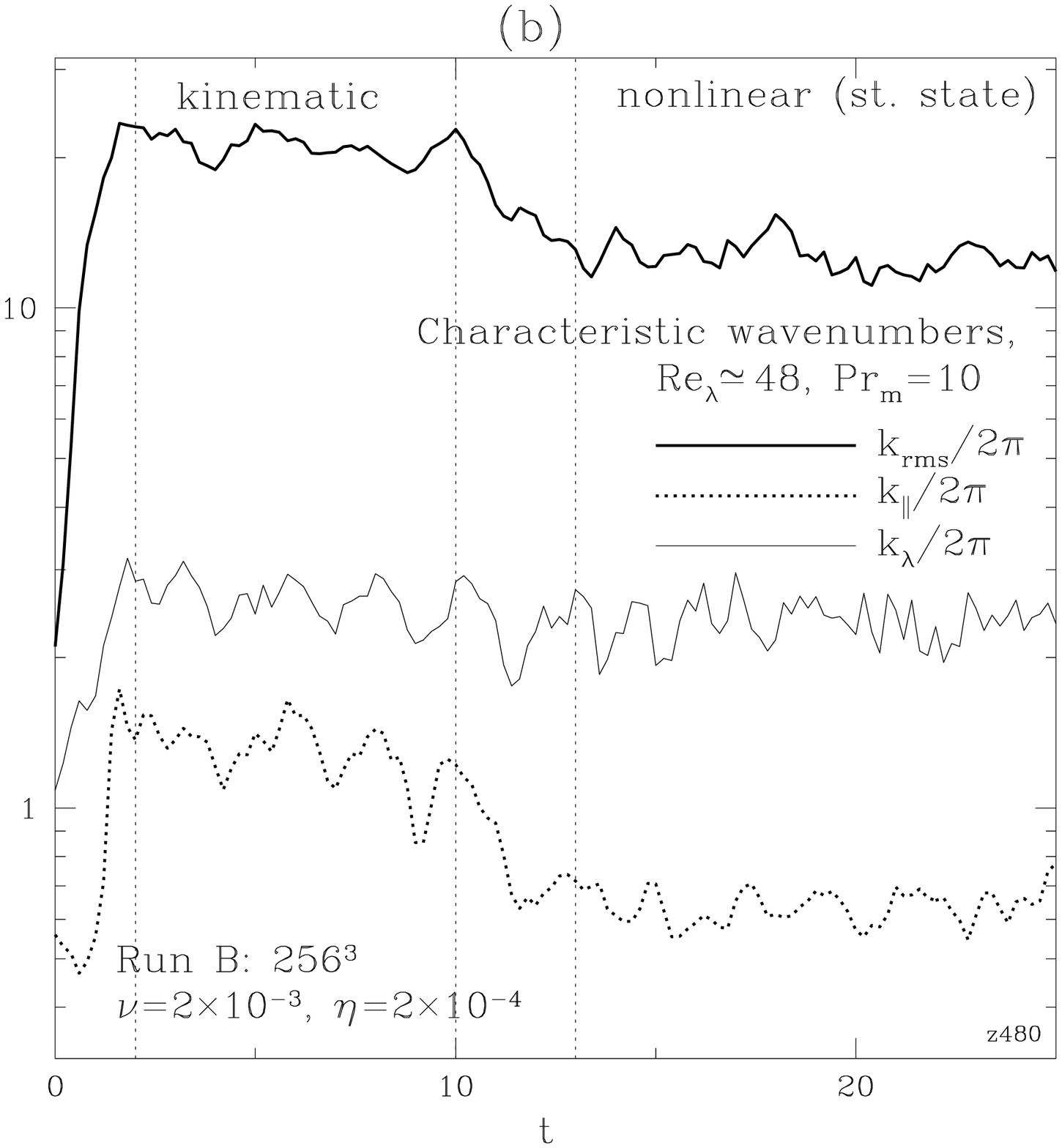}}
\caption{From kinematic dynamo to nonlinear steady state: 
(a) evolution of magnetic and kinetic energies, 
(b) evolution of characteristic wave numbers. 
Here $\kl=(\<|\nabla\vu|^2\>/\usq)^{1/2}=\sqrt{5}/\lambda$; 
$\krms$ and $\kpar$ are defined in the text.}\label{fig_hist}
\end{figure}

In order to understand what is going on, let us consider the genesis of the 
magnetic field in the isotropic MHD turbulence. As there is no mean field, 
all magnetic fields are fluctuations self-consistently 
generated by the small-scale turbulent dynamo. This type of dynamo 
is a fundamental mechanism that amplifies magnetic energy in 
sufficiently chaotic 3D flows with large enough magnetic Reynolds numbers 
(typically above~100) and~$\Pr\ge1$ 
(note that both the physics and the numerical evidence that apply 
to plasmas with $\Pr\ll 1$, e.g., stellar convective zones and 
protostellar discs, are very different: see \opencite{SCMM_lowpr}). 
The amplification is due to random stretching of the (nearly) 
frozen-in magnetic-field lines by the ambient velocity field.
What sort of magnetic fields does the dynamo make? 
During the kinematic (weak-field) stage of the dynamo,  
the growth of the field is exponential in time (\figref{fig_hist}a)
and the magnetic-energy 
spectrum is peaked at the resistive scale, $\kres\sim\Pr^{1/2}\kd$, 
and grows self-similarly (spectral profiles normalised by~$\Bsq$ 
are shown in \figref{fig_spectra}; theory is due 
to \opencite{Kazantsev}; \opencite{KA}). 
The dynamo growth rate is of the order of the turnover rate 
of the viscous-scale eddies, which are the fastest in Kolmogorov 
turbulence.

\begin{figure}[t]
\centerline{%
\begin{tabular}{c@{\hspace{1pc}}c}
\includegraphics[width=2.6in]{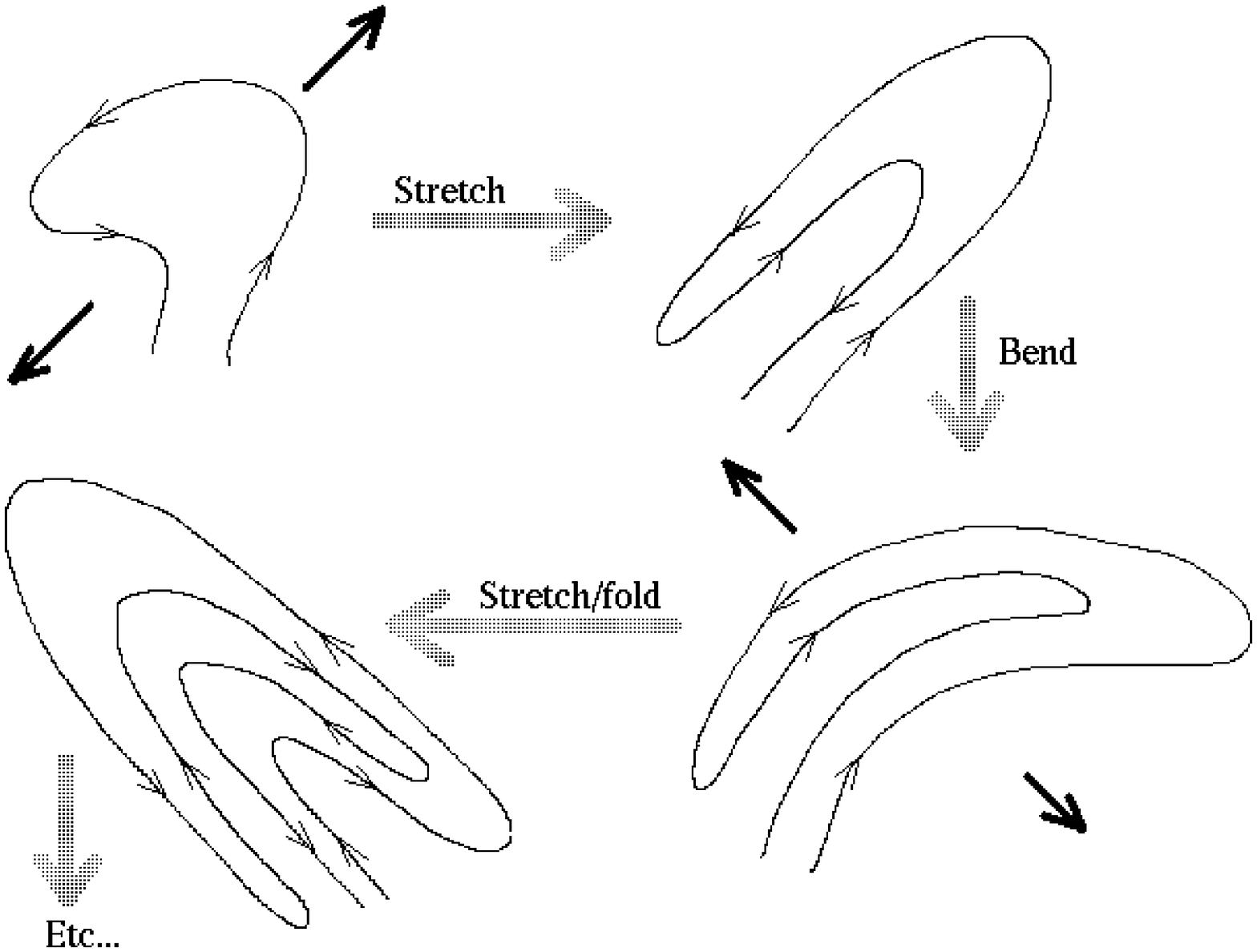} &
\includegraphics[width=1.9in]{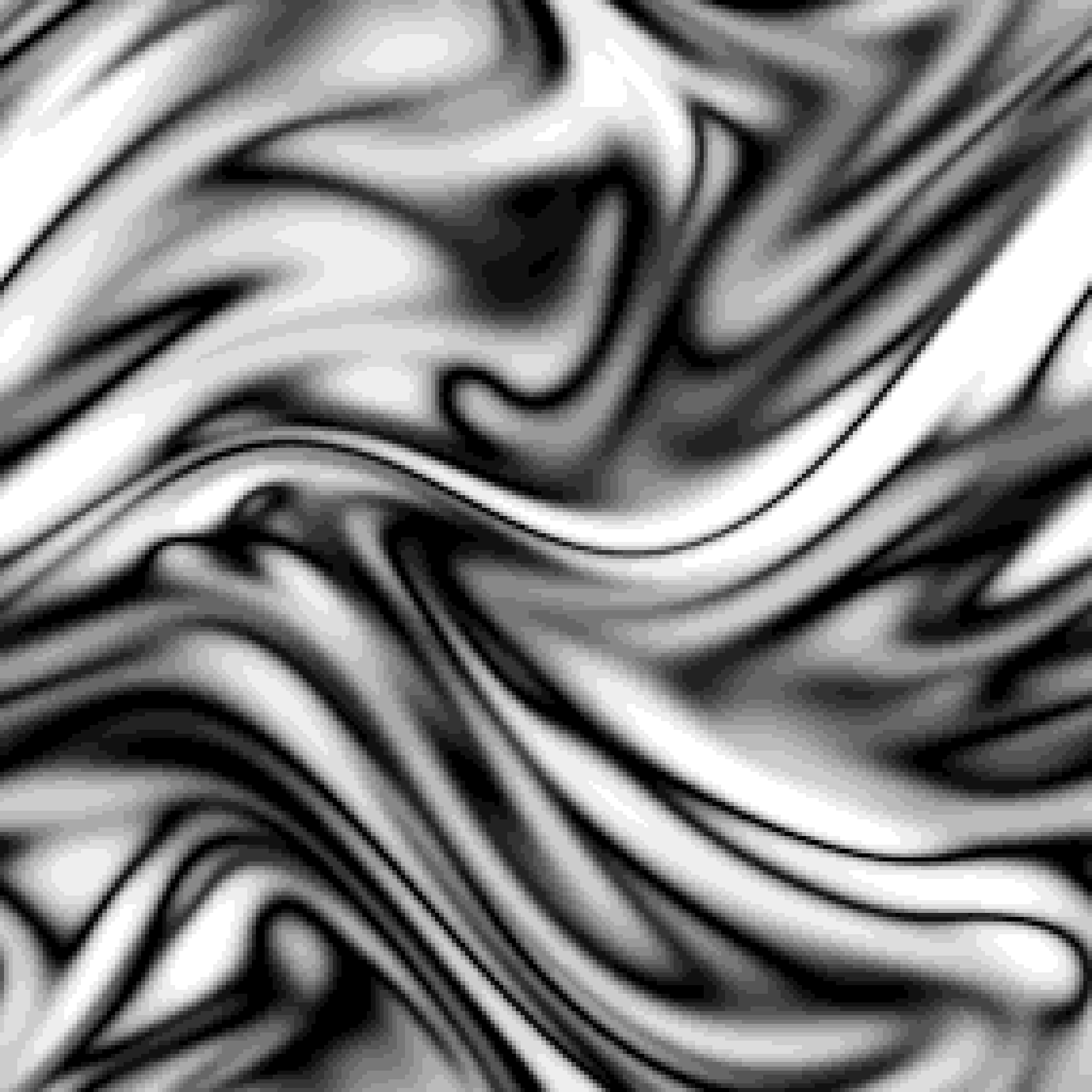} \\
(a) & (b)
\end{tabular}}
\caption{The folded fields. (a) Sketch of the stretch-and-fold 
mechanism. (b) Cross-section of field strength in a simulation 
of nonlinear dynamo ($\Pr=500$).}\label{fig_folds}
\end{figure}

Thus, the bulk of the magnetic energy 
concentrates at the resistive scale. Let us inquire, however, what actually 
happens to the field lines? It turns out that the dynamo-generated fields 
are not at all randomly tangled, but rather organised in folds 
within which the fields remain straight up to the scale of the flow 
and reverse direction at the resistive scale. \Figref{fig_folds}b 
illustrates the folded structure observed numerically. 
\Figref{fig_folds}a explains how such 
fields result from repeated application of random shear 
to the field lines. 
The folded structure is, in fact, a measurable 
statistical property of the magnetic field. A variety of diagnostics can be 
used. Ott and coworkers noticed in early 1990s that kinematically generated 
fields exhibited extreme flux cancellation due to field reversals 
(see review by \opencite{Ott_review}). Our preferred description has been 
to compare the characteristic parallel and rms wave numbers of the field 
and to study the statistics of the field-line curvature 
(\opencite{Kinney_etal}; \opencite{SCMM_folding}, \citeyear{SCTMM_stokes}). 
We have shown that (i) $\kpar=(\<|\vB\cdot\nabla\vB|^2\>/\Bfr)^{1/2}\sim\kd$ 
independently 
of~$\Pr$ while $\krms=(\<|\nabla\vB|^2\>/\Bsq)\sim\Pr^{1/2}\kd\gg\kd$ 
(\figref{fig_hist}b) (ii) the bulk of the curvature PDF 
is at flow scales. 
This confirms the picture of direction-reversing fields that are straight 
up to the scale of the flow. 
Note that the folded structure is not static: it is a net statistical effect  
of constant restretching and partial resistive annihilation 
of the magnetic fields. 

One immediate implication of the folded field structure is the criterion 
for the onset of nonlinearity: for incompressible MHD, the back reaction 
is controlled by the Lorentz tension force~$\vB\cdot\nabla\vB\sim\kpar B^2$, 
which must be comparable to other terms in the momentum equation. 
This quantity 
depends on the parallel gradient of the field and does not know about 
direction reversals. Balancing~$\vB\cdot\nabla\vB\sim\vu\cdot\nabla\vu$, 
we find that back reaction is important when magnetic energy becomes 
comparable to the energy of the viscous-scale eddies. 
Clearly, some form of nonlinear suppression of the stretching 
motions at the viscous scale must then occur. However, the eddies at 
larger scales are still more energetic than the magnetic field 
and continue to stretch it at their (slower) 
turnover rate. When the field energy reaches the energy of these 
eddies, they are also suppressed and it is the turn of yet larger 
and slower eddies to exercise dominant stretching. 
A model constructed along these lines \cite{SCHMM_ssim} leads us 
to expect a self-similar nonlinear-growth stage during which the 
magnetic energy grows~$\propto t$ and the rms wavenumber of the magnetic 
field drops~$\propto t^{-1/2}$ due to selective decay of the modes 
at the large-$k$ end of the magnetic-energy spectrum. 
The folded structure is preserved with folds elongating to 
the size~$\ls$ of the dominant stretching eddy. 
If $\Pr\gg\sqrt{\Re}$, the culmination of this process is 
a situation where total magnetic and kinetic energies 
are equal, $\Bsq\sim\usq$, 
the fields are still folded with~$\kpar\sim$~inverse box size,  
and $\krms$ has dropped by a factor of~$\Re^{1/4}$ 
compared to the kinematic case. 
Since $\Pr\gg\sqrt{\Re}$, we have~$\krms\sim\kd\Pr^{1/2}\Re^{-1/4}\gg\kd$ 
still below the viscous scale. Thus, the condition $\Pr\gg\sqrt{\Re}$ is 
necessary for $\krms$ and $\kd$ to be distinguishable in the 
nonlinear regime. It is possible that a second 
period of selective decay ensues, but this cannot be checked numerically 
because the case $\Pr\gg\sqrt{\Re}\gg1$ is unresolvable and likely 
to remain so for a long time. While astrophysical large-$\Pr$ plasmas 
are certainly in this regime, the numerics can only handle 
$\sqrt{\Re}>\Pr\gtrsim1$ (or $\Pr\gg\sqrt{\Re}\sim1$, see 
\inlinecite{SCTMM_stokes} and \figref{fig_folds}b). 
What can we learn from such numerical 
experiments? Our model predicts that the nonlinear-growth stage 
in such a situation is curtailed with $\Bsq/\usq\sim\Pr/\sqrt{\Re}<1$ 
and magnetic energy residing around the viscous scale. 
Note that all these estimates are asymptotic ones, so factors 
of order unity can obscure comparison with simulations, in which 
only very moderate scale separations can be afforded. With this 
caveat, we claim that the numerically observed state where there is 
excess magnetic 
energy at small scales but overall $\usq>\Bsq$ (\figref{fig_spectra}), 
is consistent with our prediction. The ratio $\Bsq/\usq$ does increase 
with~$\Pr$, but a conclusive parameter scan is not yet possible. 
Furthermore, the reduction of $\krms$ in the nonlinear regime, 
as well as the elongation of the folds (decrease of~$\kpar$),  
are clearly confirmed by the numerics (\figref{fig_hist}b). 

\begin{figure}[t]
\centerline{\includegraphics[width=3in]{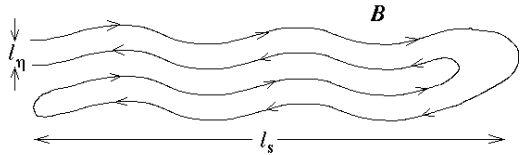}}
\caption{Alfv\'en waves superimposed on folded fields.}\label{fig_waves}
\end{figure}

Our model was based on the assumption of some effective 
nonlinear suppression 
of stretching motions. This does not 
have to mean complete suppression of all turbulence 
in the inertial range. Two kinds of motions that do not 
amplify the field can, in principle, survive. 
First, the velocity gradients 
could become locally 2D and perpendicular 
to the field (a quantitative model based on such 
two-dimensionalisation is described in 
\opencite{SCTHMM_aniso} and \opencite{SCTMM_stokes}). One should expect a large 
amount of 2D mixing of the direction-reversing field lines 
leading to very fast diffusion of the field. 
If this happens, any selective decay must be ruled out,  
i.e., the nonlinear-growth stage cannot occur 
and $\Bsq$ cannot grow above the viscous-eddy energy. 
Since numerics support selective decay and saturated values 
of~$\Bsq$ are certainly far above the energy of 
the viscous eddies, we tentatively conclude that 
mixing due to the surviving 2D motions 
is not very efficient. The second kind of allowed motions 
are Alfv\'en waves that propagate along the folded 
direction-reversing fields (\figref{fig_waves}). 
Their dispersion relation is 
$\omega_\vk=\pm(\vb\vb:\vk\vk)^{1/2}\Bsq^{1/2}$, 
where $\vb=\vB/B$ and $k$ varies between the inverse length 
of the folds ($\sim$~box size) and~$\kd$ \cite{SCHMM_ssim}. 
These waves do not stretch the field, 
do not know about the direction reversals, and 
have the same properties as standard Alfv\'en waves. 
However, Fourier transforming a field structure 
sketched in \figref{fig_waves} would not give~$\vB_\vk=\vu_\vk$ 
(scale-by-scale equipartition). Instead, the magnetic-energy 
spectrum would be heavily shifted towards small scales 
due to the presence of direction reversals. 
The Alfv\'en-wave component, while mixed up with folds in 
the magnetic field, should be manifest in the velocity 
field. We therefore expect the kinetic energy to have 
an essentially Alfv\'enic spectrum, probably~$k^{-5/3}$. 
This picture accounts for the excess small-scale 
magnetic energy observed in numerical simulations. 
Thus, {\em we conjecture that the fully developed isotropic 
MHD turbulence is a superposition of Alfv\'en waves 
and folded fields.}
The advent of truly high-resolution numerical simulations 
should make it possible to confirm or vitiate this hypothesis 
in the near future.

A more extended exposition of the ideas presented above, 
as well as a detailed account of the numerical evidence, 
are given in an upcoming paper by \opencite{SCTMM_stokes}. 

\begin{acknowledgements}
We would like to thank E.~Blackman, S.~Boldyrev, A.~Brandenburg, 
and N.~E.~Haugen for stimulating discussions. 
This work was supported by the PPARC Grant No.~PPA/G/S/2002/00075.
Simulations were done at UKAFF (Leicester) and NCSA (Illinois).
\end{acknowledgements}

\bibliographystyle{klunamed}
\bibliography{sctmm_madrid}

\end{article}
\end{document}